\documentclass[A4,prl, twocolumn, superscriptaddress ,showpacs,preprintnumbers,amsmath,amssymb,floatfix]{revtex4}
\usepackage{color}
\usepackage{graphicx}
\usepackage{dcolumn}
\usepackage{bm}
\usepackage[mathscr]{eucal}
\usepackage[nooneline]{subfigure}

\newcommand{\ud}{\mathrm{d}}

\newcommand{\bra}[1]{\langle#1\vert}
\newcommand{\ket}[1]{\vert#1\rangle}
\newcommand{\op}[2]{\ket{#1}\bra{#2}}

\newcommand{\eqn}[1]{Eq.~(\ref{#1})}
\newcommand{\eqns}[1]{Eqs.~(\ref{#1})}
\newcommand{\ignore}[1]{}

\newcommand{\commute}[2]{[#1,#2]}

\DeclareMathOperator{\Tr}{Tr}

\newcommand{\tr}[1]{\Tr\left\{#1\right\}}
\newcommand{\fig}[1]{Fig.~\ref{#1}}

\DeclareMathOperator{\ramp}{\Theta}

\newcommand{\hc}{\mathsf{H.c.}}

\newcommand{\splitting}{\phi}
\newcommand{\trel}{\tau_\mathrm{r}}
\newcommand{\tdeph}{\tau_\mathrm{d}}
\newcommand{\tmeas}{\tau_\mathrm{m}}
\newcommand{\tms}[1]{#1}
\newcommand{\sdb}[1]{#1}
\newcommand{\deletion}{}
 \usepackage{psfrag}
\newcommand{\hsys}{H_\mathrm{sys}}

\newcommand{\hL}{H_\source}
\newcommand{\hR}{H_\drain}

\newcommand{\hmeas}{H_\mathrm{meas}}
\newcommand{\hrot}{H_\mathrm{0}}
\newcommand{\hi}{H_\mathrm{I}}
\newcommand{\htot}{H_\mathrm{Tot}}

\newcommand{\hleads}{H_\mathrm{leads}}

\newcommand{\rhoi}{\rho_\mathrm{I}}
\newcommand{\rhos}{\rho}

\newcommand{\dummy}{t'}
\newcommand{\lind}{\mathcal{D}}
\newcommand{\jump}{{\mathcal J}}
\newcommand{\A}{\mathcal{A}}
\newcommand{\Tnd}{\mathcal{T}}
\newcommand{\too}{T_{00}}
\newcommand{\xoo}{\chi_{00}}

\newcommand{\Lind}{\mathcal{L}}

\newcommand{\shb}{S_{\mathrm{hb}}}

\newcommand{\slb}{S_{\mathrm{lb}}}

\newcommand{\iloc}{\bar \current_\mathrm{loc}}
\newcommand{\diloc}{\delta \current_\mathrm{loc}}

\newcommand{\source}{S}
\newcommand{\drain}{D}
\newcommand{\current}{I}
\newcommand{\evolution}{\mathcal{L}}

\begin{document}

\title{Continuous quantum measurement: inelastic tunnelling \tms{and lack of} current oscillations}

\author{T.M. Stace}\email{tms29@cam.ac.uk}
\affiliation{Cavendish Laboratory and DAMTP, University of Cambridge, Cambridge, UK}
\author{S. D. Barrett}\email{sean.barrett@hp.com}
\affiliation{Hewlett Packard Laboratories,  Filton Road, Stoke Gifford Bristol, BS34 8QZ, UK}

\date{\today}



\begin{abstract}
We study the dynamics of a charge qubit, consisting of a single electron in a double well potential coupled to a point-contact (PC) electrometer, using the quantum trajectories formalism.  Contrary to previous predictions, we show formally that, in the sub-Zeno \tms{limit},  coherent oscillations in the detector output are suppressed\tms{, and the dynamics is dominated} by inelastic processes in the PC.  Furthermore, these reduce the detector efficiency and induce relaxation even when the source-drain bias is zero. This is of practical significance since it means the detector will act as a source of decoherence. Finally, we show that the sub-Zeno dynamics \sdb{is} divided into two regimes: low- and high-bias in which the PC current power spectra show markedly different behaviour.  
\end{abstract}
\maketitle


Single shot quantum measurement of mesoscopic systems is recognised as an important goal.  Fundamentally, it will allow us to make time-resolved observations of  quantum mechanical effects in such systems, and practically it will be a necessary component in the construction of solid-state quantum information processors (QIP).  There are numerous proposals for implementing QIP in solid state systems, using doped silicon \cite{kan98}, electrostatically defined quantum dots \cite{los98} and superconducting boxes \cite{mak01}.  In these proposals, the output of the QIP is determined by measuring the position of a single electron or Cooper pair.  Single electrons hopping onto single quantum dots have been observed on a microsecond time scale using single-electron transistors (SET) \cite{lu03}. Ensemble measurements of a double well system (qubit) have been demonstrated in superconducting devices \cite{pas03}.  So far, single shot qubit measurements remain elusive.  

\begin{figure} 
\psfrag{P}{\Large$\vert 0 \rangle$}%
\includegraphics[width=5cm]{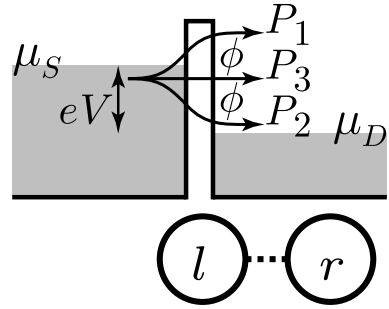}
\caption{\label{fig:schematic} Schematic of the qubit and PC showing lead energy bands.  Electrons tunnelling from the source \tms{(S)} to the drain \tms{(D)} may do so elastically or inelastically, depicted by arrows.  Different transitions induce different jumps, $P_i$, on the qubit.}
\end{figure}

It is therefore important to consider the measurement of single electron qubits by sensitive electrometers.  Two possible electrometers have been discussed to date: SETs (\deletion e.g. \cite{mak01}) and point contacts (PCs) \cite{goa01a,goa01,gur97,kor01a,kor01b,moz02,pil02,gur03,mak00}.  PCs \tms{are} sensitive charge detectors \cite{fie93,buk98,gar03,elz03}, and are the focus of this Letter.  Figure \ref{fig:schematic} illustrates the physical system we consider here, with a PC (\deletion two Fermi seas separated by a tunnel barrier) in close proximity to one of the \tms{dots}.\deletion  

\tms{Three energy scales are relevant}:  the \tms{qubit level splitting}, $\splitting$, the PC bias voltage, $e V$, and the measurement induced dephasing rate, $\Gamma_d=2(\sqrt{\current_l}-\sqrt{\current_r})^2/e$, due to the distinct currents, $\current_{l,r}$, through the PC  when the qubit electron is  held in the left ($l$) or right ($r$) well.  \tms{We ignore} environmental decoherence \cite{gur03}\deletion.  These energies define three distinct measurement regimes: 
(1) low-bias regime, \deletion $\Gamma_d/2 \ll e V<\splitting$,
(2) high-bias regime, \deletion$\Gamma_d/2 \ll \splitting <e V$ and 
(3) quantum Zeno limit, \deletion$\splitting \ll \Gamma_d/2 \ll  e V$.

 In the quantum Zeno limit, frequent weak measurements  localise the qubit, suppressing its dynamics \cite{goa01a,kor01b}. To date, little attention has been paid to the low bias regime.  Recently, the \emph{asymmetric} power spectrum was calculated for arbitrary $eV$ \cite{shn02}.  This may correspond to the emission and absorption power spectra of quanta in the \tms{PC}, as discussed in \cite{agu00}.  It has been predicted that in the limit, $\Gamma_d/2\ll \splitting$, \tms{referred}  to as the sub-Zeno limit, coherent oscillations due to the quantum dynamics of the qubit will be observed in the \deletion current through the PC \cite{goa01a,kor01a,kor01b,gur03}.  It \tms{was} also \deletion claimed that the PC is an efficient detector, \tms{i.e.}\ no information about the qubit is lost by the detector.   We present a detailed analysis of the sub-Zeno limit, showing the sharp difference in behaviour between the low- and high- bias regimes, and \deletion that previously predicted coherent oscillations are \tms{suppressed}\deletion
.  The analysis also formally yields the boundary at which approximations leading to the Zeno effect are reasonable.

For a low transparency PC, the Hamiltonians, in units where $\hbar=1$, for the qubit, leads, and their interaction are given by  \cite{goa01a,goa01,gur97,mak00,kor01a,kor01b, gur03}
\begin{eqnarray}
\hsys&=&(-\Delta\,\sigma_x-\epsilon\,\sigma_z)/2=-\splitting\,\sigma_z^{(e)}/2,\\
\hmeas&=&\sum_{k,q} (T_{k,q}+\chi_{k,q} \sigma_z) a^\dagger_{\drain,q} a_{\source,k}+\hc\\
\hleads&=&\hL+\hR=\sum_k \omega_k (a^\dagger_{\source,k} a_{\source,k}+a^\dagger_{\drain,k} a_{\drain,k})
\end{eqnarray}
where $\sigma_x=\op{l}{r}+\op{r}{l}$, $\sigma_z=\op{l}{l}-\op{r}{r}$, $\sigma_z^{(e)}=\op{g}{g}-\op{e}{e}$, $\theta=\tan^{-1}(\frac{\Delta}{\epsilon})$, $\splitting=\sqrt{\Delta^2+\epsilon^2}$, $\ket{g}=\cos(\theta/2)\ket{l}+\sin(\theta/2)\ket{r}$ and $\ket{e}=-\sin(\theta/2)\ket{l}+\cos(\theta/2)\ket{r}$.  We will adopt the convention that $\chi_{k,q}<0$ so that the left well is nearest the PC.  Furthermore we assume the PC is weakly responding so $|\xoo|\ll|\too|$.


\paragraph{Unconditional master equation.}  
\deletion The von Neumann equation for the density matrix, $R$, of the bath and qubit is
$
\dot R(t)=-i\commute{\htot}{R}.
$
To derive the master equation \tms{(ME)} for the reduced density \tms{matrix}, $\rho$, of the qubit we transform to an interaction picture with respect to the free Hamiltonian $\hrot=\hsys+\hleads$, $\hi(t)\textrm{\deletion}=e^{i \hrot t}\hmeas e^{-i \hrot t}$.  We formally integrate the von Neumann equation, then trace over lead modes \tms{\cite{gar00}},
\begin{equation}
  \dot \rhoi(t)=\Tr_{\source,\drain}\{-i\commute{\hi(t)}{R(0)}-\int_{0}^t d \dummy\commute{\hi(t)}{\commute{\hi(\dummy)}{R(\dummy)}}\},
\label{eqn:SecondOrderExpansion}
\end{equation}
where $ \hi(t)=\sum_{k,q} S_{k,q}(t)a^\dagger_{\drain,q} a_{\source,k}+\hc  $,
and $S_{k,q}(t)=e^{-i(\omega_k-\omega_q)t}(T_{k,q}+\chi_{k,q} \sigma_z(t))$.  We write $S_{k,q}(t)$  as a discrete Fourier decomposition 
$S_{k,q}(t)=e^{-i t (\omega_k-\omega_q)}(e^{-i t\splitting}P_1+e^{i t \splitting}P_2+
P_3)$,
where $P_1=P_2^\dagger=-{\xoo \sin(\theta)}\op{g}{e}$, $P_3=(\too +{\xoo \cos(\theta)}{}\sigma_z^{(e)})$.  We have assumed that $T_{k,q}=\too$ and $\chi_{k,q}=\xoo$  are constant.    

  The form of $S_{k,q}(t)$ indicates that there are three possible jump processes, indicated in \fig{fig:schematic}. $P_3$ is associated with  elastic tunnelling of electrons through the PC. $P_1$ ($P_2$) is associated with inelastic excitation (relaxation) of electrons tunnelling through the PC with an energy transfer $\phi$.  This energy is provided by the qubit which relaxes (excites) in response.  Inelastic transitions in similar systems have been described in \cite{agu00}, which calculated the current power spectrum through an open double well system due to shot noise through a nearby PC.
  
We make standard assumptions that result in a time-independent Markovian \tms{ME}.  Firstly we assume that the leads are always near thermal equilibrium so $\Tr_{\source,\drain}\{a^{(\dagger)}_{i,k}R(t)\}=0$ and $\Tr_{\source,\drain}\{a^{\dagger}_{i,k} a_{\sdb{n},k}R(t)\}=\delta_{i,j}f_i(\omega_k)\rhos(t)$, where \tms{$i,j\in\{S,D\}$}, $f_i$ is the Fermi distribution for lead $i$ and $\rhos$ is the qubit density matrix.  Secondly, if the lead correlation time is much less than other time scales, then the lower limit on the integral in \eqn{eqn:SecondOrderExpansion} can be set to $-\infty$ \cite{gar00}. 
We substitute $\sum_{k}\rightarrow\int \ud\omega_{k} g_{\source(\drain)}$ where $g_{\source(\drain)}$ are the lead density of states, which are assumed constant over the relevant energies.

Finally, we  make the rotating wave approximation (RWA), setting to zero terms in the \tms{ME}  proportional to $e^{\pm i \phi t}$.  The RWA is applicable when $\splitting\gg\nu^2 e V$ (defining the sub-Zeno limit), where $\nu=\sqrt{2\pi g_\source g_\drain} \xoo$, which is small in the low transparency, weakly responding limit. 
We also make the approximation $\rho(\dummy)\rightarrow\rho(t)$ in \eqn{eqn:SecondOrderExpansion}, which is 
 valid when $\nu^2\ll1$,
to arrive at the interaction picture \tms{ME}
\begin{eqnarray}
\dot \rhoi(t)&=&2\pi g_\source g_\drain\sum_n \left(\lind[\sqrt{\ramp(e V+\omega_n)}P_n]\rhoi(t)\right.\nonumber\\ &&\left.{}+\lind[\sqrt{\ramp(-e V-\omega_n)}P_n^\dagger]\rhoi(t)\right)\equiv \evolution_I \rho_I(t),\label{eqn:mastereqn2}
\end{eqnarray}
where $e V=\mu_\source-\mu_\drain$ is the bias applied across the PC, $\lind[B]\rho=\jump[B]\rho-\A[B]\rho$, $\jump[B]\rho=B\rho B^\dagger$, $\A[B]\rho=\frac{1}{2}(B^\dagger B \rho+\rho B^\dagger B)$, $\ramp(x)=(x+|x|)/2$  \deletion\sdb{, $\omega_1=-\omega_2=\splitting$, $\omega_3=0$} and $\mu_i$ is the chemical potential of lead $i$.  The first term in \eqn{eqn:mastereqn2} arises from forward tunnelling electrons (\tms{S-D}), whilst the second is due to reverse processes.  We note that the dynamics implicit in this \tms{ME} differs from those presented previously \cite{goa01,goa01a,gur97,kor01a,kor01b}, since it \tms{includes} inelastic jump \tms{processes}.

In the Zeno limit, $\splitting\ll\nu^2 e V$, the RWA breaks down.  Instead an alternate approximation is accurate, $e^{i\splitting t}\rightarrow1$, or equivalently $\sigma_z(t)\rightarrow\sigma_z(0)$.  In this limit, $S_{k,q}(t)=e^{-i (\omega_k-\omega_q) t}(\too+\xoo \sigma_z(0))$ has only one Fourier component, and we regain the \tms{ME} of \cite{gur97,kor01a,goa01a}. 

\paragraph{Conditional \tms{ME.} } \tms{By unravelling the \tms{ME} into evolution conditioned on measurement results, we can calculate} the observed current correlation function and the power spectrum. From the unravelling it is also possible to simulate quantum state trajectories consistent with single-shot experimental realisations of measured currents.  There is no unique unravelling for \deletion \eqn{eqn:mastereqn2}: one may add an arbitrary constant to each of the jump operators \cite{goa01} 
 in \eqn{eqn:mastereqn2}, to produce new (rescaled) operators
\begin{eqnarray}
c_1&=&\nu \sqrt{e V+\splitting}\sin(\theta)\op{g}{e}+\gamma_1,\\
c_2&=&\nu \sqrt{|e V-\splitting|}\sin(\theta)\op{e}{g}+\gamma_2,\\
c_3&=&\nu \sqrt{e V}\cos(\theta)\sigma_z^{(e)}+\Tnd\sqrt{e V},
\end{eqnarray} 
which along with a modification to the system Hamiltonian, provides an identical unconditional ME\sdb{.  Note that} $\Tnd=\sqrt{2\pi g_\source g_\drain}\too$.   

The unravelling that most accurately represents a given measurement process must be determined from physical considerations.   We discuss in detail only the unravelling in the high bias regime, in which all \deletion jumps correspond to tunnelling from \tms{source (S) to drain (D)}, as reverse processes are Pauli blocked.  Thus, when measuring currents through the leads, none of the processes are distinguished, since they all just contribute to a current.  As a result, when a jump occurs, the resultant state of the qubit is a probabilistic mixture 
$
\tilde \rho_{1c}(t+dt)=\sum_n \jump[c_n] \rho_c(t) dt
$.  
 Between jumps, it evolves smoothly according to $\dot{\tilde \rho}_{0c}(t)=-i \commute{\hsys}{\tilde\rho_{0c}(t)}-\sum_n \A[c_n]\tilde\rho_{0c}(t)$.

The constants associated with the inelastic processes, $\gamma_{1,2}$, are eliminated using energy conservation arguments.  Imagine we prepare the qubit in the state $\ket{g}$.  Then suppose, using a sensitive bolometer measuring the change in energy in the leads, we determined that an inelastic jump had followed.  We would conclude that a lead electron tunnelling from \tms{S} to \tms{D}  lost an amount of energy $-\omega_2=\phi$.  Simultaneously, the qubit evolves discontinuously,  $\ket{g}\rightarrow c_2\ket{g}$.  If $\gamma_2=0$ then this state  is $\ket{e}$, so the qubit  gains an energy $\phi$, and energy is conserved, as required.  Otherwise, if $\gamma_2\neq 0$ then it is straightforward to show that energy is not conserved on average. 
A similar argument gives $\gamma_1=0$, and these arguments also hold in the low bias regime.  
\deletion \tms{The} same reasoning applies when considering current measurements (rather than bolometric), resulting in $\gamma_1=\gamma_2=0$.

In order to determine the final constant in the unravelling, $\Tnd$, we relate the current through the PC to the jump processes according to the relation 
\begin{equation}{\current(t)} dt=e\tr{\tilde \rho_{1c}(t+dt)}
= e\sum_n \tr{\jump[c_n] \rho_c(t) } dt. \label{eqn:current}
\end{equation}
  In the absence of tunnelling ($\theta=0$), if the qubit is localised in the energy eigenstate $\ket{l}$ ($\ket{r}$) we expect a current $\current_l\equiv e T_l^2$ ($\current_r\equiv e T_r^2$) to flow through the PC.  Solving the two equations $\sum_n \tr{c_n \op{l}{l}c_n^\dagger} =e V(\Tnd+\nu)^2=T_l^2$ and $\sum_n \tr{c_n \op{r}{r}c_n^\dagger} =e V(\Tnd-\nu)^2=T_r^2 $ yields $\Tnd=({T_r+T_l})/{2\sqrt{e V}}$ and $\nu=({T_l-T_r})/{2\sqrt{e V}}$.  The average and difference  in the currents are $\iloc=({\current_r+\current_l})/{2}=e(\Tnd^2+\nu^2) e V$ and $\diloc=\current_r-\current_l=-4 e\nu\Tnd e V$. 

For general values of $\theta$  the mean current does not change, $\bar \current(\theta)=\iloc$, but the difference in current between the two localised states is
$
\delta \current(\theta)=\diloc \cos^2(\theta)
$, to lowest order in $\nu$.
Notably, the variation in current between localised states vanishes when $\theta=\pi/2$.  This demonstrates that when the energy eigenstates are completely delocalised, the PC measurement does not localise the qubit.  This is evident from the form of the jump operators, since when $\theta=\pi/2$, $c_3$ does not affect the qubit whilst $c_{1,2}$ only induce transitions between the symmetric ($\ket{g}$) and antisymmetric ($\ket{e}$) states.


\paragraph{Results.}  In the high bias regime,  the \tms{ME} is given by \eqn{eqn:mastereqn2} with $\ramp(-e V-\omega_n)=0$. 
Solving the high-bias \tms{ME}  in the interaction picture gives
\begin{eqnarray}
\rho_{ge}(t)&=&e^{-{\Gamma_d} (1+\cos^2(\theta))t/2}\rho_{ge}(0),\nonumber\\
\rho_{gg}(t)&=& \frac{\splitting+e V}{2 e V} - e^{-\Gamma_d\sin^2(\theta)t}(\frac{\splitting+e V}{2 e V}-\rho_{gg}(0)),\label{eqn:solnHB}
\end{eqnarray}
where we have defined $\Gamma_d=2\nu^2 e V$.

\begin{figure}
\includegraphics[width=8cm]{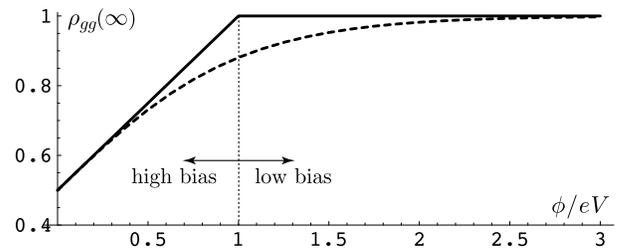}
\caption{\label{fig:QuasiThermal} Equilibrium ground state occupation probability for a qubit near  a PC (solid), and for a thermalised qubit, $\rho_{gg}^\mathrm{therm}(\infty)$, at a temperature $T=e V/2$ (dashed).}
\end{figure}

\deletion \tms{Our} analysis is valid in the low bias regime, $e V<\splitting$, \tms{also}.  The unconditional \tms{ME} is given by \eqn{eqn:mastereqn2} with $\ramp(-e V+\splitting)>0$, so it includes the term $\lind[c_2^\dagger]\rho(t)$.
\deletion 
The solution to the unconditional \tms{ME} in the low bias regime is 
\begin{eqnarray}
\rho_{ge}(t)&=&e^{-(\Gamma_d\cos^2(\theta)+\nu^2\splitting\sin^2(\theta))t}\rho_{ge}(0) ,\nonumber\\
\rho_{gg}(t)&=& 1 -e^{-2  \nu^2\splitting\sin^2(\theta)t}(1-\rho_{gg}(0)),\label{eqn:solnLB}
\end{eqnarray}
Note that at $\splitting=e V$, \eqns{eqn:solnHB} and (\ref{eqn:solnLB}) agree.

The unconditional, steady-state probability of the qubit to be found in its ground state is given by $\rho_{gg}(\infty)=\bra{g}\rho(\infty)\ket{g}$.  From \eqns{eqn:solnHB} and (\ref{eqn:solnLB}), we see that $\rho_{gg}(\infty)$ depends on the regime in which the PC is operating
\begin{equation}
\rho_{gg}(\infty)=\left\{\begin{array}{ c c}
     \frac{e V+\splitting}{2e V} & \textrm{ if }\splitting<e V \\
     1 & \textrm{ if }\splitting>e V
\end{array}\right.
\end{equation}
In contrast, if the qubit  were in thermal equilibrium with a heat bath at temperature $T$, then the ground state occupation would be $\rho_{gg}^\mathrm{therm}(\infty)={e^{\splitting/ T}}/({1+e^{\splitting/ T}})$.  Figure \ref{fig:QuasiThermal} shows these two ground state probabilities \tms{when}  $T=e V/2$. There is an evident analogy between the PC bias voltage and an external heat bath. We note that the leads are each nominally at zero temperature\deletion.

This correspondence between the PC bias and a temperature agrees with the predictions of  \cite{moz02} which shows that a PC induces effective thermal fluctuations in the position of a nearby harmonic oscillator.  Even in the limit $e V\rightarrow0$, the PC still \tms{induces relaxation}, and \deletion \eqn{eqn:solnLB} \tms{shows} that in this regime, the energy relaxation time is  $\trel^{-1}=2  \nu^2\splitting\sin^2(\theta)$, and the dephasing time is  $\tdeph^{-1}=\Gamma_d\cos^2(\theta)+\nu^2\splitting\sin^2(\theta)$.  When $e V=0$ or $\theta=\pi/2$,  $\tdeph=2\trel$, indicating that the dephasing is solely due to energy relaxation\deletion\tms{, much like the decay of an atom in a vacuum}.  This is practically significant, as it means the PC cannot be turned off merely by making $e V=0$.

\tms{Another important time scale}  is the measurement time, $\tmeas$.  Following \cite{goa01}, we calculate the initial rate at which the $z$-component of the Bloch vector increases, starting from the symmetric state $\ket{\psi(0)}=(\ket{l}+\ket{r})/\sqrt{2}$.  This is given by 
$E[\Tr\{\sigma_z^{(e)} d\rho_c(t)\}^2]/2\equiv\tmeas^{-1}dt$.  We find $\tmeas^{-1}=\Gamma_d \cos^2(\theta)+O(\nu^3\phi)$.  Also, $\tmeas\geq\tdeph$, with equality only when $\theta=0$, indicating that the detector is inefficient unless the qubit energy eigenstates are localised.


We now calculate current power spectra, using the two-time correlation function 
$
G(\tau)=E[\current(t+\tau)\current(t)]-E[\current(t+\tau)]E[\current(t)],
$
where $E[...]$ is the classical expectation.   Using \eqn{eqn:current} this can be written as \cite{goa01a}
\begin{eqnarray}
G(\tau)&=&
      e^2(\Tr\{\sum_{n,n'}\jump[c_n]e^{\Lind_S\tau}\jump[c_{n'}]\rho_\infty\}-\Tr\{\sum_{n}\jump[c_n]\rho_\infty\}^2) \nonumber\\
      &&{}+
      e^2 \Tr\{\sum_{n}\jump[c_n]\rho_\infty\} \delta(\tau),
\end{eqnarray}
where $\evolution_S$ is the Schr\"odinger picture version of the evolution operator in \eqn{eqn:mastereqn2}.   The power spectrum is then 
$
S(\omega)=2\int_{-\infty}^\infty d\tau\, G(\tau)e^{-i\omega \tau}
$.
Since $I(t)$ is the real-valued, continuously-measured current through the PC, $G(\tau)$ is manifestly symmetric.  To compute the steady state correlation function, we take the limit $t\rightarrow\infty$, so $\rho(t)\rightarrow\rho_\infty$.   
Keeping only the lowest order terms in a series expansion in $\diloc$, then taking the Fourier transform gives
\begin{equation}
\shb(\omega)=S_0-\tms{e}\,\diloc\frac{\splitting}{V}\cos(\theta)+\frac{\diloc^2\sin^2(2\theta)\Gamma_d\frac{(e V)^2-\splitting^2}{(e V)^2}}{4(\Gamma_d^2\sin^4(\theta)+ \omega^2)}\label{eqn:PSHB}\nonumber
\end{equation}
where $S_0=2e\iloc$.
%
%

\begin{figure}
\includegraphics[width=8cm]{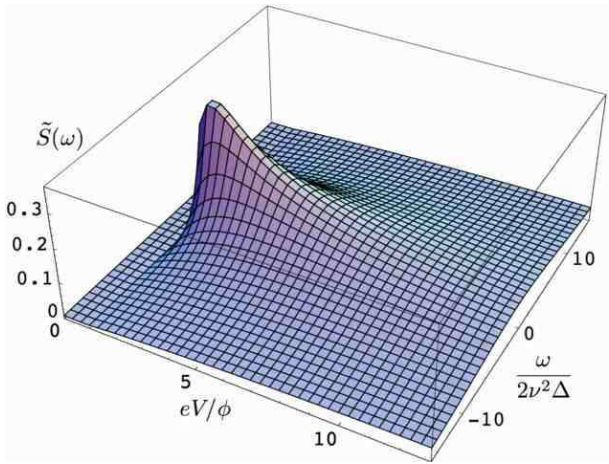}
\caption{\label{fig:PowerSpectrum} Current power spectra, $\tilde S(\omega)=\frac{2\splitting\nu^2\Delta^2}{\epsilon^2\diloc^2}(S(\omega)-S(\infty))$.  For a given $\nu$ the plot is valid in the region  $e V/\splitting\ll1/\nu^2$.  
Note the sharp change in the power spectrum at $\phi=e V$.}
\end{figure}

In the low bias regime, \deletion two distinguishable jump processes \deletion occur: \tms{S-D} electron tunnelling (\deletion elastic and inelastic), and \tms{D-S} (inelastic\deletion).  Therefore the current is related to the number of jumps in the time interval by $i\,dt=e(dN^+(t)-dN^-(t))$, where $dN^{+(-)}(t)$ counts the number of S-D (D-S) tunnelling events in the time interval $(t,t+dt)$.  
We find the steady-state, low bias current correlation function is just due to elastic tunnelling through the PC,
\begin{equation}
\slb(\omega)=S_0-e\, \diloc \cos(\theta)+O(\diloc^2).\label{eqn:PSLB}\nonumber
\end{equation}
Figure \ref{fig:PowerSpectrum} shows power spectra for different $e V/\splitting$.  These power spectra are notably different from those predicted elsewhere in the sub-Zeno limit \cite{goa01a,kor01a,kor01b,gur03}, with the absence of coherent oscillations at $\omega=\pm\splitting$.  

In conclusion, we have shown that, in the sub-Zeno \tms{limit}, inelastic tunnelling processes through a PC \tms{are important, and  coherent oscillations are absent in the detector output}, contrary to previous claims.   Projective measurements in the localised basis  are still possible as long as the energy eigenstates themselves are localised.  If the eigenstates are not localised, inelastic jumps reduce the detector efficiency.  These inelastic jumps also generate pseudo-thermal fluctuations in the qubit.   This is true even when \tms{$eV=0$}, leading to decoherence of the qubit when the detector is nominally off.  This is of practical significance in a QIP, since it means the detector will act as a source of decoherence.

TMS thanks the \tms{Hackett committee, the CVCP and Fujitsu} for financial support.  SDB acknowledges support from the E.U. NANOMAGIQC project
(Contract no. IST-2001-33186).  We thank H-S.\ Goan, W.\ J.\ Munro, T.\ Spiller, G.\ J.\ Milburn, H-A.\ Engel, R.\ Aguado, A.\ Shnirman and D.\ Averin for useful conversations.


\end{document}